\documentclass[conference,a4paper]{APSIPA2026}
\usepackage{amsmath}
\usepackage{graphicx}
\usepackage{multirow}
\usepackage{threeparttable}
\usepackage[backend=biber,style=ieee,]{biblatex}
\usepackage{geometry}
\usepackage{fancyhdr}

\fancypagestyle{firststyle}{
  \fancyhf{}
  \fancyhead[C]{2026 Asia Pacific Signal and Information Processing Association Annual Summit and Conference (APSIPA ASC)}
}

\usepackage{tikz}
\usetikzlibrary{arrows.meta,calc}
\usepackage{booktabs} 
\usepackage{adjustbox} 
\usepackage{amssymb}  
\usepackage{makecell}
\usepackage{diagbox}
\usepackage{siunitx}
\usepackage{multirow} 
\usepackage{bm}  

\usepackage{tikz}
\usetikzlibrary{positioning, fit, calc}

\usepackage{acronym}
\acrodef{DNN}{deep neural network}
\acrodef{GAN}{generative adversarial network}
\acrodef{STFT}{short-time Fourier transform }
\acrodef{TF}{time-frequency}
\acrodef{RIR}{room impulse response }
\acrodef{SNR}{signal-to-noise ratio}
\acrodef{MOS}{mean opinion score}
\acrodef{DNN}{deep neural network}
\acrodef{PDF}{probability density function}
\acrodef{FiLM}{feature-wise linear modulation }
\acrodef{TSE}{Target speaker extraction }
\acrodef{SE}{speech enhancement}
\acrodef{PF}{power law factor}
\acrodef{LR}{learning rate}
\acrodef{NAL}{noise attenuation level}
\acrodef{SOTA}{state-of-the-art}

\newcommand{\Xv}{\mathbf{X}}
\newcommand{\Sv}{\mathbf{S}}
\newcommand{\hatSv}{\widehat{\mathbf{{S}}}}
\newcommand{\Nv}{\mathbf{N}}
\newcommand{\tildeXr}{\widetilde{\mathbf{X}}_{\mathrm{r}}}
\newcommand{\tildeXi}{\widetilde{\mathbf{X}}_{\mathrm{i}}}
\newcommand{\Xr}{\mathbf{X}_{\mathrm{r}}}
\newcommand{\Xim}{\mathbf{X}_{\mathrm{i}}}
\newcommand{\tildeXm}{\widetilde{\mathbf{X}}_{\mathrm{m}}}
\newcommand{\tildeXp}{\widetilde{\mathbf{X}}_{\mathrm{p}}}
\newcommand{\barM}{\widetilde{\mathbf{M}}_\mathrm{m}}
\newcommand{\tildeYr}{\widetilde{\mathbf{Y}}_{\mathrm{r}}}
\newcommand{\tildeYi}{\widetilde{\mathbf{Y}}_{\mathrm{i}}}
\newcommand{\tildeMr}{\widetilde{\mathbf{M}}_{\mathrm{r}}}
\newcommand{\tildeMi}{\widetilde{\mathbf{M}}_{\mathrm{i}}}
\newcommand{\tildeXc}{\widetilde{\mathbf{X}}_{\mathrm{c}}}
\newcommand{\tildeYc}{\widetilde{\mathbf{Y}}_{\mathrm{c}}}
\newcommand{\fconv}{\widetilde{\mathbf{f}}_{\mathrm{conv}}}

\newcommand{\microNet}{\text{\textmu Net}}

\begin{document}

\title{\textmu Net: Ultra-Low-Memory and Low-Complexity Speech Enhancement for Embedded Digital Signal Processors}

\author{
    \authorblockN{
        Shrishti Saha Shetu\authorrefmark{1},
        Jose Miguel Martinez Aponte \authorrefmark{2}, 
        Nagashree K. S. Rao\authorrefmark{2}, 
        Sharvin Vittappan\authorrefmark{2}, \\
        Oliver Thiergart\authorrefmark{1}, 
        and Emanu\"{e}l A. P. Habets\authorrefmark{1}
    }
    \authorblockA{
        \authorrefmark{1}International Audio Laboratories, Erlangen, Germany\\ 
        \thanks{A joint institution of Fraunhofer IIS and Friedrich-Alexander-Universit{\"a}t Erlangen-N{\"u}rnberg (FAU), Germany.}
        \authorrefmark{2}Fraunhofer IIS, Erlangen, Germany
    }
 
}

\maketitle
\def\thefootnote{}\footnote{ * A joint institution of Fraunhofer IIS and Friedrich-Alexander-Universit{\"a}t Erlangen-N{\"u}rnberg (FAU), Germany.}\addtocounter{footnote}{-1}\def\thefootnote{\arabic{footnote}}

\thispagestyle{firststyle}
\pagestyle{empty}

\vspace{-1em}
\begin{abstract}
  Speech enhancement on embedded digital signal processors (DSPs) imposes strict constraints on memory footprint, computational complexity, latency, and support for integer operations. Although recent DNN-based approaches have addressed these challenges individually, no unified framework in the literature simultaneously addresses all these requirements for practical deployment. In this work, we propose \textmu Net, an ultra-low-memory, low-complexity, and low-latency end-to-end DNN model. The proposed method requires only $90$~KB of static memory and $28$~MMACs, while supporting an algorithmic latency as low as $4$~ms with performance comparable to state-of-the-art methods of similar complexity. Our experiments demonstrate that \textmu Net is compatible with neural accelerators and supports full integer-arithmetic operations on consumer DSP platforms such as Cadence Tensilica HiFi 4/5.
\end{abstract}


\section{Introduction}
\label{sec:intro}

Speech enhancement aims to suppress background noise while preserving intelligibility and improving the perceived quality of the speech signal. In recent years, \ac{DNN}-based techniques have risen to prominence, demonstrating substantial performance improvements over conventional statistical signal processing methods~\cite{boll1979suppression, ephraim2003speech, Valin2020APA, choi2021real, hu2020dccrn, braun2021towards, schroter2022deepfilternet2}. However, most \ac{SOTA} \ac{DNN} models involve significant computational complexity and large memory footprints. Furthermore, they typically operate in higher-latency regimes (e.g., $10$--$40$~ms), which,  while suitable for general telecommunications~\cite{vary2023digital}, are often infeasible for hearable and wearable applications. Applications such as hearing aids, live dialogue enhancement for multimedia content, and hearables in transparency mode require very-low-latency processing~\cite{stone2003tolerable, amazon, denk2020acoustic}. These applications typically rely on algorithmic execution on resource-constrained consumer embedded DSPs, such as Cadence Tensilica HiFi 4/5, which often strictly constrain processing to integer operations.

Several recent approaches have proposed novel architectures to address computational efficiency, latency, and memory constraints~\cite{schroter2022deepfilternet, shetu2023ultra, rong2024gtcrn, larraza2026fast, wu2025ultra, cheng2025modulating}. In~\cite{rong2024gtcrn}, an equivalent rectangular bandwidth (ERB) filter bank is used for feature extraction, and computational overhead is reduced through grouped convolutions and RNNs~\cite{ma2018shufflenet}. Conversely, ULCNet~\cite{shetu2023ultra} operates in the \ac{STFT} domain, achieving efficiency via channel-wise feature reorientation~\cite{liu2020channel} and a two-stage processing framework for magnitude mask estimation and phase correction. While various low-latency DNN techniques have been investigated in literature, including asymmetric analysis-synthesis window pairs, learnable transforms, trainable filterbank equalizers, and future frame prediction~\cite{wu2025ultra, luo2019conv, zheng2022low, wang2022stft}; these often rely on computationally heavy models. In many practical scenarios, low latency requirements are coupled with even stricter computational limits; this joint optimization of latency, memory, and fixed-point (int8) operations support remains insufficiently addressed in the literature.

In this work, inspired by the ULCNet architecture~\cite{shetu2023ultra}, we propose \microNet \footnotemark, a fully quantizable, lightweight DNN model for speech enhancement applications. Our primary contributions are as follows:
\begin{itemize}
    \item We propose a novel ultra-low-memory, low-complexity architecture that supports algorithmic latencies down to $4$~ms and achieves performance comparable to \ac{SOTA} methods of similar complexity.
    \item We demonstrate that the proposed \microNet\ is fully quantizable to int8 and suitable for deployment on embedded DSPs with neural accelerator support.
    \item We incorporate a configurable noise attenuation control mechanism that allows users to trade off background noise suppression against speech quality, enabling adaptation to diverse acoustic scenarios.
\end{itemize}

\footnotetext{ Several aspects of the proposed \microNet~are protected by patent applications.}

\section{Proposed Methods}
\label{sec:PM}

\subsection{Input Pre-processing}
\label{sec:InPre}

We assume an additive signal model in the \ac{STFT}-domain where \(\Xv\), \(\Sv\), and \(\Nv\) denote the noisy signal, the clean speech signal, and the noise components, respectively.  In our work, the input features provided to the DNN are the magnitude and phase features computed from the modified power law compressed \cite{shetu2023ultra} real and imaginary parts of the noisy signal $\Xv$  with a \ac{PF} of $\alpha$, as follows:
\begin{align}
&\tildeXr = \mathrm{sign}(\Xr) \odot |\Xr|^{\alpha} ;   &\tildeXi = \mathrm{sign}(\Xim) \odot |\Xim|^{\alpha}
\label{eq:powerlaw}
\end{align}
where $\odot$ denotes the Hadamard product, and the $\mathrm{sign}(\Xv_{\mathrm{r/i}})$ denotes that the original sign of  real and imaginary part $\Xv_{\mathrm{r/i}}$  of $\Xv$ is retained for $\widetilde{\Xv}_{\mathrm{r/i}}$. Subsequently, the power law compressed magnitude spectrogram and phase component of the noisy signal are obtained as follows:
\begin{align}
&\tildeXm = \sqrt{\tildeXr^2+ \tildeXi^2} ; &\tildeXp = \arctan\left({\frac{\tildeXi}{\tildeXr} }\right)
\label{eq:Input Feature}
\end{align}

\begin{figure}[t]
\centering
\begin{tikzpicture}[
node distance = 4mm and 6mm, 
>=stealth,
line cap=round,
line join=round,
block/.style={
    draw,
    rectangle,
    minimum width=2.2cm,    
    minimum height=6.5mm,   
    align=center,
    fill=white,
    font=\scriptsize,       
    line width=0.5pt
},
stage1/.style={draw=red, line width=0.6pt, inner sep=4pt, rounded corners},
stage2/.style={draw=blue, line width=0.6pt, inner sep=6pt, rounded corners},
arrow/.style={->, thick},
dasharrow/.style={->, thick, dash dot}
]

\node[font=\scriptsize] (noisy) { \textbf{Noisy Signal} $\Xv$};

\node[block, below=of noisy] (pre) {Input Preprocessing};
\node[block, below=of pre,fill=gray!20] (cs1) {C-SubFR + C-SamFR};
\node[block, below=of cs1] (conv) {Conv Block};
\node[block, below=of conv] (pw1) {Pointwise Conv};
\node[block, below=of pw1] (split) {Subband Splitting};
\node[block, below=of split,fill=gray!20] (gru) {Shared Subband GRU};

\node[block, right=of pre,fill=gray!20] (csub2) {C-SubFR};
\node[block, below=of csub2] (inter) {Intermediate\\ Feature Computation};
\node[block, below=of inter] (concat1) {Feature Concat.};
\node[block, below=of concat1,fill=gray!20] (linear) {Shared Linear\\ Projection};
\node[block, below=of linear] (concat2) {Feature Concat.};

\node[block, right=of csub2] (cnn) {CNN};
\node[block, below=of cnn] (pw2) {Pointwise Conv};
\node[block, below=of pw2] (clean) {CRM-Multplication};
\node[block, below=of clean] (power) {Power Law\\ Decompression};

\node[below=6mm of power, font=\scriptsize] (out) {\textbf{Clean Estimate} $\hatSv$};

\draw[arrow] (noisy) -- (pre);

\draw[arrow] (pre) -- node[right, font=\scriptsize] {$\tildeXm$} (cs1);
\draw[arrow] (cs1) -- node[right, font=\scriptsize] {$\tildeXc$} (conv);
\draw[arrow] (conv) -- node[right, font=\scriptsize] {$\fconv$} (pw1);
\draw[arrow] (pw1) --  (split);
\draw[arrow] (split) -- (gru);

\coordinate (gruout) at ($(gru.east)+(3mm,0)$);
\draw[-] (gru.east) --  (gruout);
\draw[arrow] (gruout) |- (concat2.west);

\draw[arrow] (concat2) -- (linear);
\draw[arrow] (linear) -- (concat1);
\draw[arrow] (concat1) -- node[right, font=\scriptsize] {$\barM$} (inter);
\draw[arrow] (inter) -- node[right, font=\scriptsize] {$\{\tildeYr,\tildeYi\}$} (csub2);

\draw[arrow] (csub2) -- (cnn);
\draw[arrow] (cnn) -- (pw2);
\draw[arrow] (pw2) -- node[right, font=\scriptsize] {$\tildeMr,\tildeMi$} (clean);
\draw[arrow] (clean) -- node[right, font=\scriptsize] {$\widetilde{\mathbf{S}}$} (power);
\draw[arrow] (power) -- (out);

\draw[dasharrow]
(pre.east) -- ++(0.2,0)
|- node[pos=0.25,right,font=\scriptsize] {$\tildeXp$}
(inter.west);

\coordinate (xrxi_start)  at ($(pre.north)+(0.8cm,0mm)$); 
\coordinate (xrxi_up)     at ($(xrxi_start)+(0,5mm)$);          
\coordinate (xrxi_across) at ($(xrxi_up)+(3.4cm,0)$); 
\coordinate (xrxi_down)   at ($(clean.west)+(-3.7mm, 0mm)$);
\coordinate (xrxi_in)     at (clean.west);

\draw[dasharrow]
(xrxi_start) -- (xrxi_up)
-- (xrxi_across)
-- (xrxi_down)
-- (xrxi_in);

\node at ($(xrxi_up)!0.5!(xrxi_across)+(0,1.5mm)$) [font=\scriptsize] {$\{\tildeXr, \tildeXi\}$};

\node[stage1, fit=(noisy)(pre)(gru)(concat2)(csub2),
label={[red, font=\bfseries\scriptsize]below:First Stage}] {};

\node[stage2, fit=(cnn)(out),
label={[blue, font=\bfseries\scriptsize]below:Second Stage}] {};

\end{tikzpicture}
\caption{Proposed $\microNet$ Architecture. The processing blocks highlighted with a ``gray'' background indicate the modifications introduced in this work from the original ULCNet model~\cite{shetu2023ultra}. }
\label{fig:unet}
\end{figure}
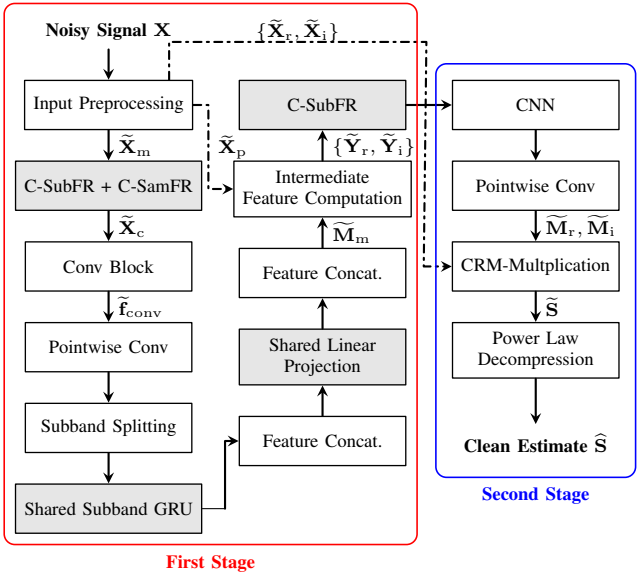

\subsection{\textmu Net Architecture}
\label{sec:unet}

In this work, as shown in Fig.~\ref{fig:unet}, we propose \microNet, which utilizes a two-stage backbone inspired by the ULCNet architecture~\cite{shetu2023ultra} with several key structural modifications. While ULCNet was originally designed for speech enhancement, it has subsequently proven effective across various related tasks, including acoustic echo cancellation, dereverberation, and beamforming~\cite{shetu2025align, shetu2024hybrid, rao2025low, bologni2026two}. In the proposed \microNet\ architecture, the first stage estimates a magnitude mask $\barM \in [0,1]$, while the second stage refines the intermediate features to estimate a final complex ratio mask (CRM).

\noindent \textbf{Channel-wise Feature Reorientation:} The power-law compressed magnitude features $\tildeXm$ are first subjected to a hybrid feature reorientation process to capture both local and global spectral dependencies. We employ a hybrid approach combining the channel-wise subband feature reorientation \linebreak (C-SubFR) method \cite{liu2020channel, shetu2023ultra} and the channel-wise sampling-based feature reorientation (C-SamFR) method \cite{shetu2025align}. To emphasize the perceptually significant low-frequency bands, we extract $2$ subbands, each spanning $43$ frequency bins. For the C-SamFR stage, a sampling factor of $6$ is applied across non-contiguous frequency bins, yielding $6$ sub-sampled feature sets of dimension $43$. This results in a reoriented feature set $\widetilde{\mathbf{X}}_c \in \mathbb{R}^{B \times T \times F \times C}$, where $B$, $T$, $F = 43$, and $C = 8$ denote the batch size, time frames, frequency bins, and channel dimensions, respectively.  For brevity, $B$ and $T$ are omitted in subsequent descriptions.

\noindent \textbf{Convolutional and Recurrent Processing:} 
The reoriented features are processed through a convolutional (Conv) block comprising four layers, each having $32$ filters with a kernel size of $(1, 3)$. Frequency-axis downsampling is achieved via strided convolutions with a factor of $2$ in the final three layers. Each convolution is followed by batch normalization and ReLU activation. The resulting feature map, $\fconv \in \mathbb{R}^{32 \times 6}$, is refined by a $1 \times 1$ point-wise convolution with $24$ filters and flattened into a $144$-dimensional vector. In this work, we employ standard convolutions rather than depthwise separable convolutions (as originally proposed in \cite{shetu2023ultra}) to optimize implementation for embedded hardware and ensure quantization support. While depthwise separable convolutions reduce the theoretical parameter count, they often suffer from poor hardware utilization on consumer DSPs due to fragmented memory access patterns. Furthermore, to learn temporal dependencies with parameter efficiency, the flattened features are split into two subbands and processed by a shared GRU with $64$ hidden units. This weight-sharing strategy allows the model to learn common temporal dynamics across subbands while significantly reducing the total parameter count, resulting in a $128$-dimensional latent feature vector $\mathbf{h}$.

\noindent \textbf{Shared Linear Projection:} 
The $128$-dimensional latent features $\mathbf{h}$ are processed by a shared linear projection block to estimate an intermediate real magnitude mask. To ensure spectral consistency during upsampling, we implement an overlapping sliding window approach. The latent feature vector $\mathbf{h}$ is partitioned into four segments $\mathbf{h}_k$ with \linebreak $k \in \{1,2,3,4\}$ of length $40$, defined by the index ranges $[0, 40]$, $[24, 64]$, $[56, 96]$, and $[88, 128]$. Each segment is projected through a shared linear layer:
\begin{equation}
    \mathbf{m}_k = \sigma(\mathbf{W} \, \mathbf{h}_k + \mathbf{b}), 
\end{equation}
where $\mathbf{W} \in \mathbb{R}^{64 \times 40}$ represents the shared weight matrix, $\mathbf{b}$ is a shared bias vector and $\sigma(\cdot)$ denotes the sigmoid activation function. The resulting projections $\mathbf{m}_k \in \mathbb{R}^{64 \times 1}$ are concatenated to estimate the final real-valued mask $\barM \in \mathbb{R}^{256 \times 1}$. This shared-weight mechanism further reduces the parameter count while ensuring that the learned filters are invariant across different segments of the feature space.

\noindent \textbf{Intermediate Feature Computation:} 
As a prerequisite for the second stage, the estimated magnitude mask $\barM$ is combined with the noisy phase $\tildeXp$ to compute intermediate features. The real and imaginary components are given by
\begin{align}
&\tildeYr = \barM \odot \cos{\tildeXp} ; &\tildeYi =  \barM \odot \sin{\tildeXp}
\label{eq:IntermediateFeature}
\end{align}
Subsequently, $\tildeYr$ and $\tildeYi$ are concatenated along the channel dimension. To further reduce the computational complexity of the second stage, these features are processed using the C-SubFR method, yielding a feature set $\tildeYc \in \mathbb{R}^{64\times 8}$.

\noindent \textbf{Clean Speech Estimation:} 
The second stage utilizes a CNN block comprising two convolutional layers, each with $32$ filters, followed by a $1 \times 1$ point-wise convolution with $8$ output channels. The clean speech is estimated by applying a CRM-multiplication \cite{hu2020dccrn} and by applying power-law decompression \cite{shetu2023ultra, shetu2025align}.

\subsection{Loss Functions}
\label{sec:LF}
To evaluate the proposed $\microNet$, we employ three distinct loss functions: Mean Squared Error ($\mathcal{L}_{\text{MSE}}$), Multi-Scale ($\mathcal{L}_{\text{MS}}$), and Multi-Target ($\mathcal{L}_{\text{MT}}$). Following \cite{shetu2023ultra}, $\mathcal{L}_{\text{MSE}}$ is computed in the compressed frequency domain for aggressive noise suppression. On the contrary, the MS loss combines a time-domain cosine similarity (CS) loss with a frequency-domain MSE, i.e.,
\begin{equation*}
\vspace{-.1cm}
\mathcal{L}_\text{MS} = \sum_j \frac{1}{K} \sum_{k=1}^{K} \text{CS}\left(\mathbf{s}_{jk}, \hat{\mathbf{s}}_{jk}\right)+ \underbrace{\sum_i \left\| |\mathbf{S}_i|^{\alpha} - |\widehat{\mathbf{S}}_i|^{\alpha} \right\|_{\text{F}}^2}_{\mathcal{L}_{\text{spec}}},
\vspace{-.1cm}
\end{equation*}
where $j=\{1,2,\ldots,K\}$ indexes the segment lengths from the set 
$\mathcal{L} = \{16, \dots, 128\}$\,ms. The index $i \in \{1, 2, \dots, I\}$  represents the STFT window sizes from $\mathcal{W} = \{16,\dots, 64\}$\,ms \cite{choi2021real} with $\left\| \text{.}\right\|_\text{F}$ denoting the Frobenius norm, and $\mathbf{S}_i = \ac{STFT}_i(\mathbf{s})$ is the $i$-th STFT.
 
Finally, the MT loss $\mathcal{L}_\text{MT}$ is a modified version of the frequency-domain MS loss, where, in addition to comparing magnitudes, a phase component is introduced:
\begin{equation*}
\vspace{-.1cm}
\mathcal{L}_\text{MT} = \mathcal{L}_\text{spec} + \sum_{i} \left\| \left| \mathbf{S}_i \right|^{\alpha}\odot e^{j\bm{\phi}_\mathbf{S}} - \left| \widehat{\mathbf{S}}_i \right|^{\alpha}\odot e^{j\bm{\phi}_{\widehat{\mathbf{S}}}} \right\|_\text{F}^2,
\vspace{-.1cm}
\end{equation*}
where $\bm{\phi_\mathbf{S}}$ and $\bm{\phi_{\widehat{\mathbf{S}}}}$ denote the phase component of $\mathbf{S}$ and $\widehat{\mathbf{S}}$, respectively.\\

\subsection{Noise Attenuation Control}
\label{sec:NAC}
We employ a post-processing method for user-defined control of the \ac{NAL} of our algorithm and trade-off with speech quality, inspired by \cite{braun2015residual}. Given the enhanced estimate $\mathbf{\hat{s}}$ and the estimated residual noise $\mathbf{\hat{n}} = \mathbf{x} - \mathbf{\hat{s}}$ in  time domain, the post-processed output at different \ac{NAL} (in dB) is formulated as
\begin{equation}
    \mathbf{\hat{s}}_{-\text{dB}} = \mathbf{\hat{s}} + \beta \; \mathbf{\hat{n}},
\end{equation}
\vspace{-0.2cm}
where 
\begin{equation}
    \beta= \sqrt{\frac{P_{\hat{s}}}{P_{\hat{n}} \cdot 10^{(\text{NAL}_{\text{dB}} / 10)}}}   
\end{equation}
is the scaling factor determined by a target attenuation level $\text{NAL}_{\text{dB}}$ to maintain a controlled noise floor, with $P_{\hat{s}}$ and $P_{\hat{n}}$ denoting the mean power of the enhanced speech and residual noise, respectively.

\section{Experiment and Results}
\label{ER}
\subsection{Training Details}
\label{sec:TD}
In our experiments, we used the Interspeech 2020 DNS challenge dataset \cite{reddy2020interspeech}, and 
noisy mixtures were generated at a $16$ kHz sampling rate
with \ac{SNR} ranges between $-10$ dB to
$30$ dB. In total, we created a training dataset of around 1000 hrs. In $50\%$ of the training and validation dataset, we convolved clean speech
with a \ac{RIR} randomly selected from
the \ac{RIR} set provided in \cite{reddy2020interspeech}. For \ac{STFT}, we use a squared-root Hann window with a $32$ms window length. During training, we used the Adam optimizer with an initial \ac{LR} of $4 \times 10^{-4}$ and a scheduler that reduces the \ac{LR} by a factor of $10$ every $3$ epochs. For data augmentation, we incorporated random low-pass filtering, upsampling, and different STFT windows. Unless explicitly specified, we use \ac{PF} as $0.3$ in all our experiments.

\begin{table}[t]
\centering
\caption{Model complexity and objective results on the DNS Challenge dataset. The subscript for different $\microNet$ variations indicates either the loss functions or different $\text{NAL}_{\text{dB}}$.} 
\resizebox{\columnwidth}{!}{
\begin{tabular}{lccccc}
\toprule
\textbf{Method} & \shortstack{\textbf{Params} \\ \textbf{(K)}} & \textbf{MMACs} & \textbf{PESQ} & \textbf{SI-SDR} & \shortstack{\textbf{BAK} \\ (\textbf{MOS})} \\
\midrule
Noisy & -- & -- & 1.58 & 9.07 & 2.62 \\
\midrule
RNNNoise \cite{valin2018hybrid} & 60 & 40  & 2.04 & 12.66 & 3.95 \\
GTCRN \footnotemark \cite{rong2024gtcrn}    & 48 & 33  & 2.26 & 14.62 & 3.98\\
\midrule
$\microNet$  V2  & 52 & 32  & 1.81 & 12.28 & 3.92 \\
$\microNet$  V3  & 55 & 32  & 1.85 & 12.44 & 3.96 \\
\midrule
$\microNet_{\text{MSE}}$  & \multirow{5}{*}{\textbf{46}} & \multirow{5}{*}{\textbf{28}} & 1.90 & 13.24 & \textbf{4.03} \\
$\microNet_{\text{MT}}$   &  &  & 2.18 & 12.74 & 3.95 \\
$\microNet_{\text{MS}}$  &  &  & 2.13 & 13.27 & 3.99 \\
$\microNet_{-25 \text{dB}}$  &  &  & 2.24 & \textbf{13.61} & 3.55 \\
$\microNet_{-30 \text{dB}}$  &  &  & \textbf{2.27} & 13.53 & 3.71 \\
\bottomrule
\end{tabular}
}
\label{tab:complexity_results}
\end{table}

\footnotetext{ The effective complexity of GTCRN is even higher, as it requires buffering up to $16$ frames of intermediate features and hence not comparable to the frame-by-frame processing supported by RNNoise and \microNet.}

\begin{figure}[t]
\centering
    \input{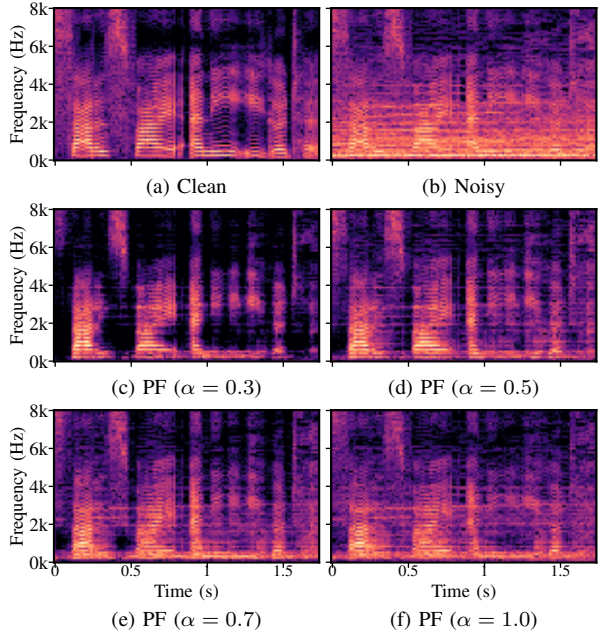}
\vspace{-1em}
  \caption{Spectrograms of (a) a clean speech signal and (b) noisy signal, enhanced by our proposed $\microNet_{\text{MSE}}$ with different power-law factor: (c)-(f).}
\label{fig:spectograms}
\end{figure}

\subsection{Comparison with Baseline Methods}
\label{sec:comp}

In this work, we used RNNoise~\cite{valin2018hybrid} and GTCRN~\cite{rong2024gtcrn} as the primary baseline methods. We also developed two additional baselines based on \microNet, where we replaced the shared subband GRU with shared subband causal self-attention~\cite{vaswani2017attention} (referred to as \microNet\ V2) and shared gated convolution~\cite{oord2018parallel} (referred to as \microNet\ V3). We utilize PESQ~\cite{rix2001perceptual}, SI-SDR~\cite{le2019sdr}, and BAK (MOS) (from DNSMOS)~\cite{reddy2022dnsmos} as objective evaluation metrics. The objective results presented in Tab.~\ref{tab:complexity_results} on the DNS-challenge non-reverberant test dataset~\cite{reddy2020interspeech} demonstrate that our proposed \microNet, trained on various loss functions and with different $\text{NAL}_{\text{dB}}$, achieves comparable performance to the baseline methods with significantly lower complexity and parameter counts\footnote{Demo: \url{https://sshetu-iis.github.io/uNet/ulm/}}. Specifically, \microNet\ trained on MSE loss with an \ac{PF} of $\alpha=0.3$ outperforms the baselines in terms of noise suppression, achieving a BAK (MOS) of $4.03$. This aggressive noise suppression behavior for this configuration was previously reported in~\cite{shetu2023ultra} and is visualized in Fig.~\ref{fig:spectograms}.

\begin{figure}[t]
\centering
\includegraphics[width=0.95\linewidth]{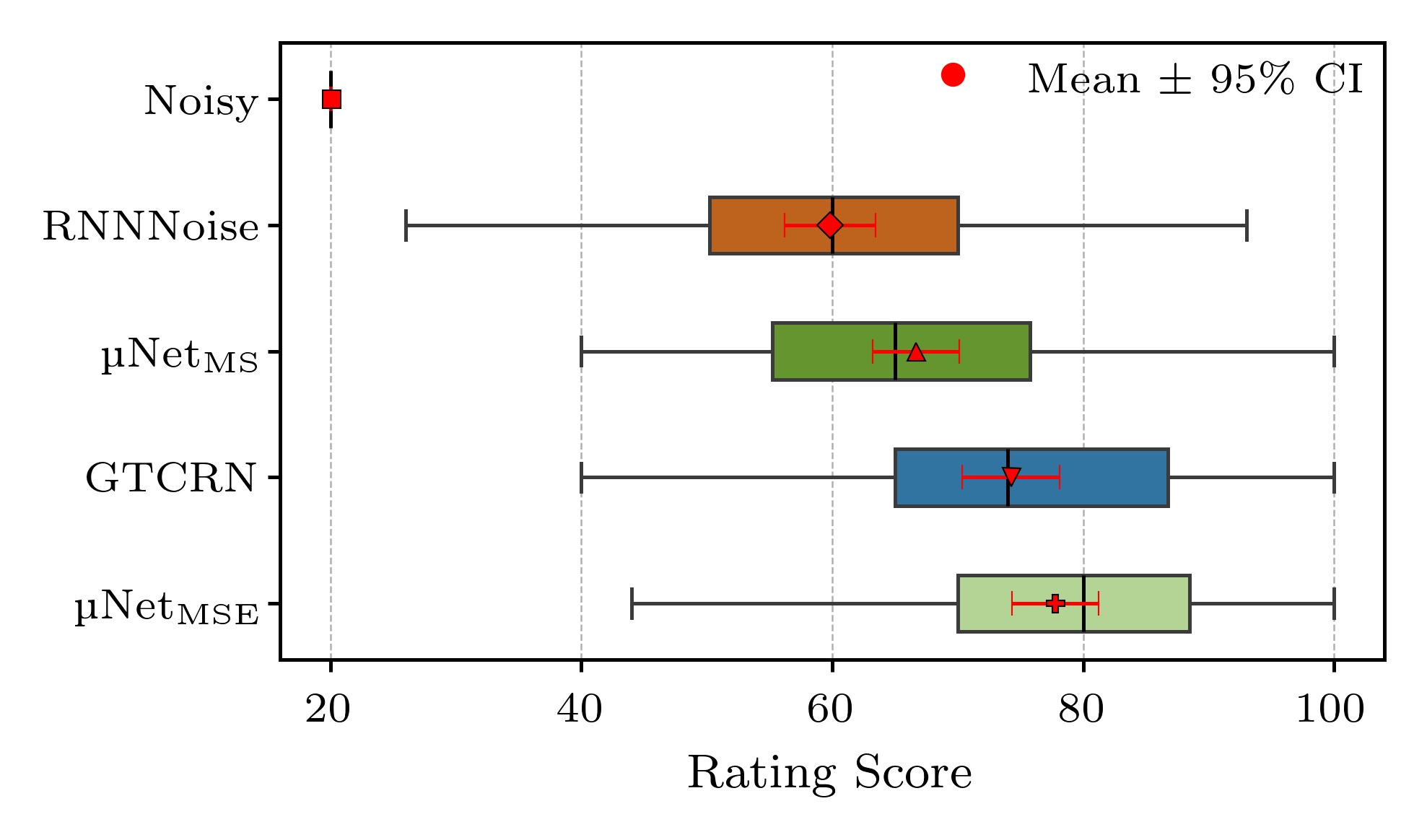}
\vspace{-1em}
\caption{Multi-stimuli listening test results with the float32 models at $16$~ms latency. The $95\%$ confidence interval is shown as red whiskers.} 
\label{fig:Lis1}
\vspace{-1.5em}
\end{figure}

As evident from Fig.~\ref{fig:spectograms}, this over-suppression can lead to distortion of the speech signal by suppressing non-harmonic components; however, this effect can be mitigated by utilizing the proposed noise attenuation control, adjusting the \ac{PF} ($\alpha$), or employing different loss functions. The proposed $\microNet_{\text{MSE}}$   with a $\text{NAL}_{\text{dB}}$ of $-30$~dB achieves a PESQ of $2.27$, outperforming the baselines in PESQ improvement. Additionally, \microNet\ variants trained on MS and MT loss functions achieve comparable results. The GTCRN outperforms the competeing methods in terms SI-SDR improvement, and achieves the highest SI-SDR of $14.62$~dB. Furthermore, we conducted a multi-stimulus listening test with $10$ listeners using the webMUSHRA framework~\cite{schoeffler2018webmushra} with $10$ randomly selected samples, following~\cite{shetu2023ultra}. The results, as shown in Fig.~\ref{fig:Lis1}, depict that our proposed \microNet\ achieves the highest mean MUSHRA score of $77.78$, outperforming GTCRN ($74.24$). Notably, the results indicate that \microNet$_{\text{MSE}}$ is perceptually preferred by users despite its over-suppression characteristics.

\begin{figure}[t]
\centering
\includegraphics[width=0.95\linewidth]{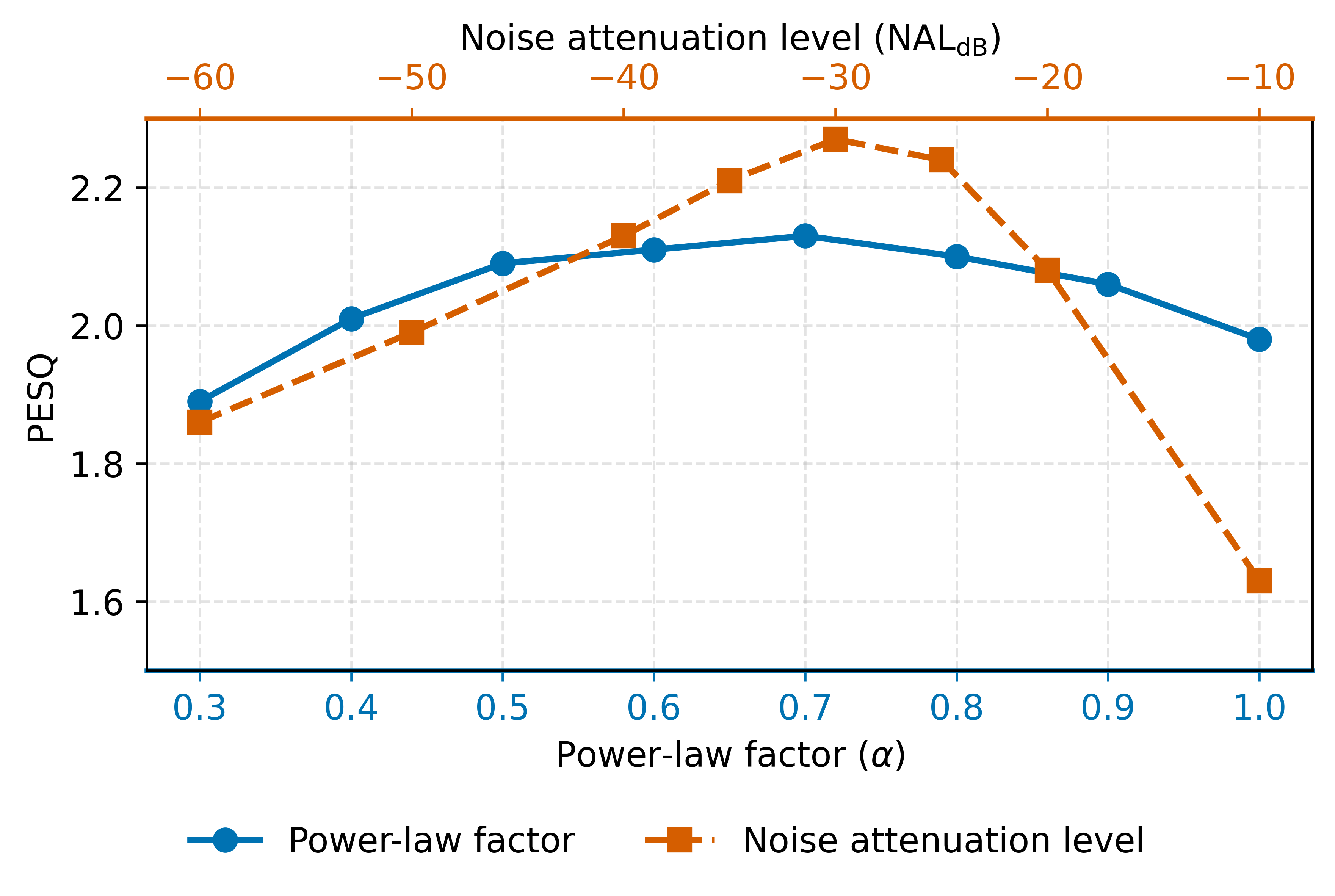}
\vspace{-1em}
\caption{Relationship between different \ac{PF} ($\alpha$) and $\text{NAL}_{\text{dB}}$ using \microNet$_{\text{MSE}}$ model. }
\label{fig:PFNAL}
\vspace{-1em}
\end{figure}

\subsection{Power Law Factor vs Noise Attenuation Level}
\label{sec:PNAL}
We further investigate the relationship between the \ac{PF} and various \ac{NAL} settings. The results, shown in Fig.~\ref{fig:PFNAL}, demonstrate that increasing the \ac{PF} for the proposed \microNet$_{\text{MSE}}$ helps maintain better speech quality, as evident by improved PESQ scores. However, as previously noted, this improvement comes at the cost of reduced noise suppression, which is functionally equivalent to setting a higher \ac{NAL}. In our experiments, we observed that while most listeners prefer aggressive noise suppression, many are highly sensitive to speech distortions, which is also reported in~\cite{sugiyama2022user}. Consequently, in real-world consumer devices and applications, it is essential to manage this trade-off~\cite{ochiai2024rethinking}. In this context, the proposed noise attenuation control with \microNet$_{\text{MSE}}$ is shown to be highly effective; as \microNet\ can effectively suppress the noise and hence as observed, for $\text{NAL}_{\text{dB}}$ up to $-35$~dB, the speech quality improves. This provides users with an intuitive mechanism to choose appropriate configurations for different acoustic scenarios.

\begin{table}[!t]
\centering
\caption{Objective performance comparison of \microNet\ across varying algorithmic latencies for float32 and int8 quantized models on the DNS Non-Reverberant dataset.}

\begin{tabular}{c cc cc}
\toprule
\textbf{Latency} & \multicolumn{2}{c}{\textbf{float32}} & \multicolumn{2}{c}{\textbf{int8}} \\
\cmidrule(lr){2-3} \cmidrule(lr){4-5}
 &  $\Delta$PESQ & $\Delta$SI-SDR & $\Delta$PESQ & $\Delta$SI-SDR \\
\midrule
16 ms   & 0.35 & 3.59 & 0.40 & 3.55 \\
08 ms  & 0.34 & 3.09 & 0.24& 2.31\\
04 ms  & 0.21 & 2.52 & 0.10 & 0.50\\
\bottomrule
\end{tabular}
\label{tab:objective_latency_first}
\vspace{-1.5em}
\end{table}

\subsection{Low Latency and Quantization}
\label{sec:LQ}
We further experimented with different latencies for \microNet. As our proposed method runs purely frame-by-frame, the overall latency is primarily determined by the \ac{STFT} synthesis window length used in the overlap-add algorithm. In our experiments, we use an asymmetric window pair, similar to those proposed in~\cite{wang2021deep,mauler2007low}, consisting of a Hann window for analysis and a shorter Hann window for synthesis. The results in Tab.~\ref{tab:objective_latency_first} show that \microNet\ can be optimized for latencies as low as $4$~ms without significant degradation. Specifically, \microNet\ achieves a $0.21$ PESQ and $2.52$~dB SI-SDR improvement on the DNS non-reverberant test set with algorithomic latency of $4$~ms. We also quantized \microNet\ using the TensorFlow TFLite post-training framework~\cite{david2021tensorflow} to fully int8. We used $10$ curated samples from the training dataset for tuning quantization variables. Our results indicate that at $16$~ms latency, int8 quantization has no adverse effect on performance; however, performance degrades as latency decreases. This degradation may be attributed to the drifting of GRU states due to more frequent updates in low-latency configurations~\cite{larraza2026fast}. 

\begin{figure}[!t]
\centering
\includegraphics[width=0.95\linewidth]{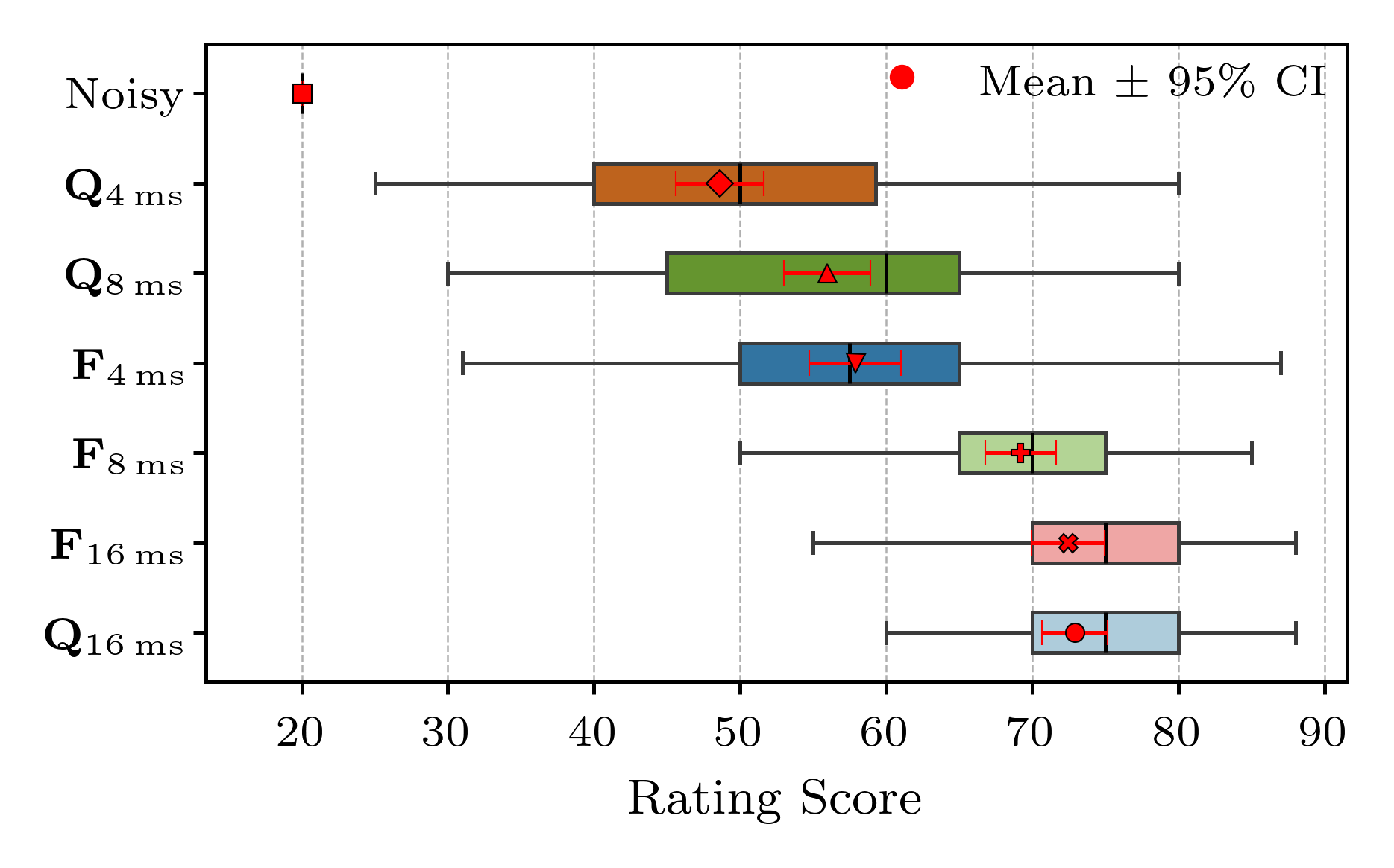}
\vspace{-1em}
\caption{
Multi-stimulus listening test results for float32 (F) and int8-quantized (Q) \microNet$_{\text{MSE}}$ models. Subscripts indicate algorithmic latencies; "F" and "Q" represent 32-bit floating-point and 8-bit integer quantization, respectively}
\label{fig:lis2}
\vspace{-1.5em}
\end{figure}

Furthermore, we conducted another multi-stimulus listening test with the same setup as previously described to evaluate the perceptual effects of lower latency and quantization. The results, shown in Fig.~\ref{fig:lis2}, depict that for $16$~ms latency, quantization has no adverse perceptual impact. In terms of latency, at $4$~ms we observe a sharp decline in user preference. The $4$~ms model achieves a mean MUSHRA score of $57.87$, compared to mean scores of $69.20$ and $72.44$ for the $8$~ms and $16$~ms methods, respectively. Notably, even though objective results indicate only minimal performance improvement for the int8-quantized $4$~ms \microNet, listeners still prefer the enhanced output over the noisy signal.

\def\thefootnote{}\footnote{ 
The authors gratefully acknowledge the scientific support and HPC resources provided by the Erlangen National High Performance Computing Center (NHR@FAU) of the FAU under the NHR project b262dc18@csnhr.nhr.fau.de. NHR funding is provided by federal and Bavarian state authorities.}\addtocounter{footnote}{-1}\def\thefootnote{\arabic{footnote}}

\subsection{DSP and Sampling Rate Support}
\label{sec:DSP}
Our proposed \microNet\ requires only $90$~KB of static memory, corresponding to the memory required to store the model parameters and weights on the DSPs (excluding additional platform-dependent dynamic workspace required during inference). \microNet\ is fully compatible with various embedded DSPs, namely ARM Cortex M, ADI SHARC, Qualcomm Hexagon and Cadence Tensilica HiFi 4/5. Notably, \microNet\ runs in real-time on Cadence Tensilica HiFi 4  on NXP RT685 DSP with a cycles requirement of $70$ MHz. Regarding sampling rates, while this work focuses on $16$ kHz, \microNet\ can be easily configured for higher sampling rates. This is achieved by adjusting the parameterization for the channel-wise feature reorientation methods, the shared subband GRU, and the shared linear projection block. 

\section{Conclusions}
We proposed \microNet\ for speech enhancement applications, which requires only $90$\,KB of static memory and $28$\,MMACs, supports algorithmic latencies as low as 4\,ms, and is fully quantizable to int8 for deployment on neural-accelerator-equipped platforms such as Airoha AB159x. We demonstrated that these constraints can be met without sacrificing competitive performance, as evidenced by objective evaluations on the DNS Challenge dataset and perceptual listening tests. Furthermore, we employed a configurable noise attenuation control mechanism that allows users to balance noise suppression aggressiveness against speech quality, making \microNet\ adaptable to diverse acoustic scenarios and consumer device requirements. Future work could explore extending \microNet\ with quantization-aware training strategies to further close the performance gap at ultra-low latencies.

\printbibliography

@string{icassp_short = "Proc. IEEE Int. Conf. Acoust., Speech, Signal Process."}

@string{iwaenc_short = "Proc. Int. Workshop Acoust. Signal Enhanc."}

@string{taslp_short = "IEEE/ACM Trans. Audio, Speech, Language Process."}

@string{isca= "Proc. INTERSPEECH"}

@string{icml_short = "Proc. Int. Conf. Mach. Learn."}

@string{ open_res = "J. Open Res. Softw."}

@string{ eusipco = "Proc. Eur. Signal Process. Conf."}

@string{ mmsp = "Proc. IEEE Int. Workshop Multimedia Signal Process. "}

@string{ neurisp_short = "Proc. Adv. Neural Inf. Process. Syst. "}

@inproceedings{le2019sdr,
  title={{SDR}--half-baked or well done?},
author = {Le Roux, J. and Wisdom, S. and Erdogan, H. and Hershey, J. R.},
  booktitle=icassp_short,
  year={2019}
}

@inproceedings{rix2001perceptual,
  title={Perceptual evaluation of speech quality ({PESQ})-a new method for speech quality assessment of telephone networks and codecs},
author = {Rix, A. W. and Beerends, J. G. and Hollier, M. P. and Hekstra, A. P.}, 
booktitle=icassp_short,
  year={2001}
}

@inproceedings{shetu2023ultra,
  title={Ultra Low Complexity Deep Learning Based Noise Suppression},
  author={Shetu, S. S. and Chakrabarty, S. and Thiergart, O. and Mabande, E.},
  booktitle= icassp_short,
  year={2024}
}

@article{schoeffler2018webmushra,
  title={{webMUSHRA}—A comprehensive framework for web-based listening tests},
author = {M. Schoeffler et al.},
  journal=open_res ,
  volume={6},
  year={2018}
}

@inproceedings{liu2020channel,
  title={Channel-wise subband input for better voice and accompaniment separation on high resolution music},
  author={Liu, H. and Xie, L. and Wu, J. and Yang, G.},
  booktitle=isca,
  year={2020}
}

@inproceedings{shetu2025align,
  title={{Align-ULCNet}: Towards low-complexity and robust acoustic echo and noise reduction},
  author={Shetu, S. S. and Desiraju, N. K. and Mack, W. and Habets, E. A. P. },
  booktitle=eusipco,
  year={2025},
}

@inproceedings{hu2020dccrn,
  title={{DCCRN}: Deep complex convolution recurrent network for phase-aware speech enhancement},
  author={Y. Hu  et al.},
  booktitle=isca,
  year={2021}
}

@inproceedings{choi2021real,
  title={Real-time denoising and dereverberation with tiny recurrent {U-Net}},
  author={H.-S. Choi  et al.},
  booktitle=icassp_short,
  year={2021},
}

@inproceedings{valin2018hybrid,
  title={A hybrid {DSP}/deep learning approach to real-time full-band speech enhancement},
  author={Valin, J.-M.},
  booktitle=mmsp,
  year={2018},
}

@inproceedings{rong2024gtcrn,
  title={{GTCRN}: A speech enhancement model requiring ultralow computational resources},
  author={X. Rong et al.},
  booktitle=icassp_short,
  year={2024},
}

@inproceedings{oord2018parallel,
  title={Parallel wavenet: Fast high-fidelity speech synthesis},
  author={A. Oord  et al.},
  booktitle=icml_short,
  year={2018},
}

@inproceedings{vaswani2017attention,
  title={Attention is all you need},
  author={Vaswani, A. and others},
  booktitle=neurisp_short,
  year={2017}
}

@inproceedings{reddy2022dnsmos,
  title={{DNSMOS P. 835}: A non-intrusive perceptual objective speech quality metric to evaluate noise suppressors},
author = {Reddy, C. K. A. and Gopal, V. and Cutler, R.},

  booktitle=icassp_short,
  year={2022},
}

@inproceedings{reddy2020interspeech,
  title     = {The {Interspeech} 2020 Deep Noise Suppression Challenge: Datasets, Subjective Testing Framework, and Challenge Results},
  author    = {C. K. A. Reddy et al.},
  booktitle = isca,
  year      = {2020}
}

@inproceedings{sugiyama2022user,
  title={User preference between residual noise and speech distortion in speech enhancement},
  author={Sugiyama, A. and Shimada, O. and Nomura, T.},
  booktitle=iwaenc_short,
year={2022},
}

@inproceedings{wang2021deep,
  title={Deep neural network based low-latency speech separation with asymmetric analysis-synthesis window pair},
  author={Wang, S. and Naithani, G. and Politis, A. and Virtanen, T.},
  booktitle=eusipco,
  year={2021}
}

@inproceedings{mauler2007low,
  title={A low delay, variable resolution, perfect reconstruction spectral analysis-synthesis system for speech enhancement},
  author={Mauler, D. and Martin, R.},
  booktitle=eusipco,
  year={2007},
}

@inproceedings{david2021tensorflow,
  title={{TensorFlow} lite micro: Embedded machine learning for {TinyML} systems},
  author={R. David et al.},
  booktitle={Proc. Mach. Learn. Syst.},
  year={2021}
}

@inproceedings{larraza2026fast,
  title={Fast-{ULCNet}: A fast and ultra low complexity network for single-channel speech enhancement},
  author={Larraza, N.A. and de Koeijer, N.},
  booktitle=icassp_short,
  year={2026}
}

@inproceedings{braun2021towards,
  title={Towards efficient models for real-time deep noise suppression},
  author={Braun, S. and Gamper, H. and Reddy, C. K. and Tashev, I.},
  booktitle=icassp_short,
  year={2021},
}

@inproceedings{schroter2022deepfilternet2,
  title={{DeepFilterNet2}: Towards real-time speech enhancement on embedded devices for full-band audio},
  author={Schr{\"o}ter, H. and Maier, A. and Escalante-B, A.N. and Rosenkranz, T.},
  booktitle=iwaenc_short,
  year={2022},
}

@inproceedings{Valin2020APA,
  title={A Perceptually-Motivated Approach for Low-Complexity, Real-Time Enhancement of Fullband Speech},
  author={Valin, J.-M.},
  booktitle=isca,
  year={2020},
}

@book{vary2023digital,
  title={Digital Speech Transmission and Enhancement},
  author={Vary, P. and Martin, R.},
  year={2023},
  publisher={John Wiley \& Sons}
}

@article{stone2003tolerable,
  title={Tolerable hearing aid delays. III. {Effects} on speech production and perception of across-frequency variation in delay},
  author={Stone, M. A. and Moore, B. C.},
  journal={Ear and Hearing},
  year={2003},
}

@misc{amazon,
  title={Dialogue Boost: How {Amazon} is using {AI} to enhance {TV} and movie dialogue},
  author={Amazon},
  journal={Amazon Science
},
  year={2025},
}

@article{denk2020acoustic,
  title={Acoustic transparency in hearables—technical evaluation},
  author={Denk, F. and Schepker, H. and Doclo, S. and Kollmeier, B.},
  journal={J. Audio Eng. Soc.},

  year={2020},
}

@article{boll1979suppression,
  title={Suppression of acoustic noise in speech using spectral subtraction},
  author={Boll, S.},
  journal=taslp_short,
  year={1979},
}

@article{ephraim2003speech,
  title={Speech enhancement using a minimum-mean square error short-time spectral amplitude estimator},
  author={Ephraim, Y. and Malah, D.},
  journal=taslp_short,
  year={2003},
}

@article{ochiai2024rethinking,
  title={Rethinking processing distortions: Disentangling the impact of speech enhancement errors on speech recognition performance},
  author={Ochiai, T. and Iwamoto, K. and Delcroix, M. and Ikeshita, R. and Sato, H. and Araki, S. and Katagiri, S.},
  journal=taslp_short,
  year={2024},
  publisher={IEEE}
}

@inproceedings{schroter2022deepfilternet,
  title={{DeepFilterNet}: A low complexity speech enhancement framework for full-band audio based on deep filtering},
  author={Schroter, H. and Escalante-B, A. N. and Rosenkranz, T. and Maier, A.},
  booktitle=icassp_short,
  year={2022},
}

@inproceedings{wu2025ultra,
  title={Ultra-low latency speech enhancement-a comprehensive study},
  author={Wu, H. and Braun, S.},
  booktitle=icassp_short,
  year={2025},
}

@inproceedings{cheng2025modulating,
  title={Modulating state space model with slowfast framework for compute-efficient ultra low-latency speech enhancement},
  author={L. Cheng et al.},
  booktitle=icassp_short,
  year={2025},

}

@inproceedings{ma2018shufflenet,
  title={Shufflenet v2: Practical guidelines for efficient {CNN} architecture design},
  author={Ma, N. and Zhang, X. and Zheng, H.-T. and Sun, J.},
  booktitle={Proc. Eur. Conf. Comput. Vis. },
  year={2018}
}

@article{luo2019conv,
  title={{Conv-TasNet}: Surpassing ideal time--frequency magnitude masking for speech separation},
  author={Luo, Y. and Mesgarani, N.},
  journal=taslp_short,
  year={2019},
}

@article{zheng2022low,
  title={Low-latency monaural speech enhancement with deep filter-bank equalizer},
  author={C. Zheng et al.},
  journal={J. Acoust. Soc. Am.},
  year={2022},
}

@article{wang2022stft,
  title={{STFT}-domain neural speech enhancement with very low algorithmic latency},
  author={Wang, Z.-Q. and Wichern, G. and Watanabe, S. and Le Roux, J.},
  journal=taslp_short,
  year={2022},
}

@inproceedings{shetu2024hybrid,
  title={A hybrid approach for low-complexity joint acoustic echo and noise reduction},
  author={Shetu, S.S. and Desiraju, N. K. and Aponte, J. M. M. and Habets,  E. A. P. and Mabande, E.},
  booktitle=iwaenc_short,
  year={2024},
}

@inproceedings{rao2025low,
  title={Low-complexity neural speech dereverberation with adaptive target control},
  author={Rao, N.K.S and Chetupalli, S. R. and Shetu, S. S. and Habets,  E. A. P. and Thiergart, O.},
  booktitle=icassp_short,
  year={2025},
}

@article{bologni2026two,
  title={A two-step approach for speech enhancement in low-{SNR} scenarios using cyclostationary beamforming and {DNNs}},
  author={Bologni, G. and Larraza, N. A. and Heusdens, R. and Hendriks, R. C.},
  journal={arXiv preprint arXiv:2602.12986},
  year={2026}
}

@inproceedings{braun2015residual,
  title={Residual noise control using a parametric multichannel Wiener filter},
  author={Braun, S. and Kowalczyk, K. and Habets, E. A. P.},
  booktitle=icassp_short,
  year={2015}
}

\end{document}